\documentclass[twocolumn,aps,pra,floatfix,showpacs]{revtex4}
\usepackage{amsmath}
\usepackage{amssymb}
\usepackage{graphicx}
\usepackage{pst-all,pstricks-add}

\newcommand{\ket}[1]{\lvert #1 \rangle}           
\newcommand{\bra}[1]{\langle #1 \lvert}           
\newcommand{\expv}[1]{\langle #1 \rangle}         
\newcommand{\innerprod}[2]{\left< #1 \vert #2 \right>}

\newcommand{\cbd}[4]{\sum_{k=#2}^{#1}\binom{#1}{k} \left(#3\right)^{k} \left(#4\right)^{#1-k}}  

\newcommand{\eps}{\varepsilon}                 
\newcommand{\epseff}{\eps_\mathrm{eff}}
\newcommand{\epscrit}{\eps_\mathrm{c}}
\newcommand{\epscritone}{\eps_\mathrm{c\; necc}}
\newcommand{\epscritall}{\eps_\mathrm{c\; suff}}

\newcommand{\psallq}[3]{p_\mathrm{success\; all}^{\phantom{success\! all} \mathrm{q}}(#1,#2,#3)} 
\newcommand{\pfallq}[3]{p_\mathrm{fail\; all}^{\phantom{fail\! all} \mathrm{q}}(#1,#2,#3)} 
\newcommand{\pfoneq}[3]{p_\mathrm{fail\; one}^{\phantom{fail\! one} \mathrm{q}}(#1,#2,#3)} 
\newcommand{\psoneq}[3]{p_\mathrm{success\; one}^{\phantom{success\! one} \mathrm{q}}(#1,#2,#3)} 
\newcommand{\pfallc}[2]{p_\mathrm{fail\; all}^{\phantom{fail\! all} \mathrm{c}}(#1,#2)} 
\newcommand{\psallc}[2]{p_\mathrm{success\; all}^{\phantom{success\! all} \mathrm{c}}(#1,#2)} 
\newcommand{\Mmax}{M_\mathrm{max}}                              
\newcommand{\Mmin}{M_\mathrm{min}}                              

\newcommand{\effpn}[0]{\frac{1+\epseff}{2}}                
\newcommand{\effmn}[0]{\frac{1-\epseff}{2}}                
 
\newcommand{\U}[0]{\hat{U}_{\mathrm{alg}}}

\DeclareMathOperator{\Trace}{Tr}
\DeclareMathOperator{\nint}{nint}
\DeclareMathOperator{\order}{O}

\begin{document}

\author{Tomasz M.~Kott}
\affiliation{Department of Physics, University of Maryland, College Park, MD 20742 USA}
\email{tkott@umd.edu}

\author{David Collins}
\affiliation{Department of Physical and Environmental Sciences, Mesa State College, 1100 North Avenue, Grand Junction, CO 81501, USA}
\email{dacollin@mesastate.edu}
\homepage{http://home.mesastate.edu/~dacollin/index.html}
\thanks{Author to whom correspondence should be addressed.}

\title{Statistical comparison of ensemble implementations of Grover's search algorithm to classical sequential searches.}

\begin{abstract}
We compare pseudopure state ensemble implementations, quantified by their initial polarization and ensemble size, of Grover's search algorithm to probabilistic classical sequential search algorithms in terms of their success and failure probabilities. We propose a criterion for quantifying the resources used by the ensemble implementation via the aggregate number of oracle invocations across the entire ensemble and use this as a basis for comparison with classical search algorithms. We determine bounds for a critical polarization such that the ensemble algorithm succeeds with a greater probability than the probabilistic classical sequential search. Our results indicate that the critical polarization scales as $N^{-1/4}$ where $N$ is the database size and that for typical room temperature solution state NMR, the polarization is such that the ensemble implementation of Grover's algorithm would be advantageous for $N \gtrsim 10^{22}.$
\end{abstract}


\maketitle

\section{Introduction}
\label{sec:intro}

Conventional implementations of quantum algorithms entail application of unitary transformations and projective measurements to a single, multi-qubit quantum system which is initially prepared in a known pure state~\cite{nielsen00,lebellac06,ekert96,steane98}. In the alternative ensemble paradigm, of which solution state nuclear magnetic resonance (NMR) is the most prominent example, the algorithm is implemented on an ensemble of multiple  identical, noninteracting quantum systems~\cite{chuang98a,cory98,cory97,gershenfeld97,marx00,vdsypen01,vdsypen00,vdsypen04,negrevergne05,negrevergne06}. 

The principal differences between the two paradigms manifest themselves in the preparation, measurement and post-measurement stages and arise from: (i) the impossibility of isolating any given ensemble members and applying preparation, evolution and measurement operations to these alone while ignoring the remaining ensemble members and (ii) the fact that available initial states are mixed. The principal similarity between the paradigms is that it is possible to apply identical unitary evolution operators to each ensemble member at any stage of the algorithm.  A variety of schemes have been developed to deal with the initialization, measurement and algorithm output issues in ensemble implementations of quantum algorithms~\cite{chuang98a,cory97,schulman99,chuang98b,knill98a,collins02}. Collectively these issues are resolved via the use of a  \emph{pseudopure initial state}, for whose preparation there are several known schemes~\cite{schulman99,cory98,chuang98b,chuang98a,knill98a}, described by the density operator
\begin{equation}
  \hat{\rho}_i=\frac{\left(1-\eps\right)}{2^n}\, \hat{I}^{\otimes n}+ \eps\, \ket{\psi_i}\bra{\psi_i}
  \label{eq:pseudopure}
\end{equation}
where $n$ is the number of qubits, $\ket{\psi_i}$ is the initial pure state required by the conventional (single quantum system) implementation of the algorithm and $0 \leqslant \eps \leqslant 1$ is called the \emph{polarization.} Under the unitary transformation for the algorithm, $\U$, the density operator transforms to 
\begin{align}
 \hat{\rho}_\mathrm{final} & =\frac{\left(1-\eps\right)}{2^{n}}\, \hat{I}^{\otimes n}
                             + \eps \, \U \ket{\psi_{i}}\bra{\psi_{i}}\U^{\dagger} \nonumber \\
                           & =\frac{\left(1-\eps\right)}{2^{n}}\, \hat{I}^{\otimes n}
                             + \eps \, \ket{\psi_\mathrm{final}}\bra{\psi_\mathrm{final}}
 \label{eq:pseudopurefinal}
\end{align}
where
\begin{equation}
  \ket{\psi_\mathrm{final}} := \U \ket{\psi_{i}} 
\end{equation}
is the final, pre-measurement state of the quantum system in the conventional implementation of the algorithm. This gives the appearance of a conventional realization of the algorithm on a subset of the ensemble which is initially described by the density operator $\ket{\psi_i}\bra{\psi_i},$ which corresponds to a pure state. In most proposals for and realizations of ensemble implementations the algorithm output is extracted from the expectation values of single qubit traceless observables~\cite{cory97,knill98a,chuang98a,collins02}, as we describe in Sec.~\ref{sec:groverensemble}. Such expectation values are deterministic. However, in real implementations the number of ensemble members is finite and expectation values can only be approximated by sample averages. Thus, regardless of the nature of the original quantum algorithm, ensemble implementations are probabilistic (except possibly in the special case where $\eps=1$ which reduces to a conventional implementation). Although this basic feature has been noted~\cite{cory97,knill98}, there has been limited discussion of its implications with regard to reasonable comparisons of ensemble implementations of quantum algorithms to probabilistic classical competitors. Such issues have been addressed in the context of various ensemble implementations of single bit output algorithms such as the Deutsch-Jozsa algorithm~\cite{arvind03,anderson05}, resulting in bounds of the polarization beneath which a classical probabilistic algorithm will succeed with greater probability than the ensemble implementation of the quantum algorithm. 

In this article we compare the performance of an ensemble implementation of the Grover search algorithm to probabilistic classical sequential searches. Our aim is to determine the polarization as a function of both ensemble and database size such that the ensemble implementation of the quantum algorithm succeeds with greater probability than its classical competitor. The primary distinction between this and previous work on the Deutsch-Jozsa algorithm is that the output of the Grover algorithm is determined from measurements on many qubits and that these measurement outcomes are correlated.

The article is organized as follows. We describe the conventional Grover search algorithm in Sec.~\ref{sec:groversingle}. In Sec.~\ref{sec:groverensemble} we describe an ensemble implementation of the Grover search algorithm and explicitly provide a protocol for extracting an algorithm output. The bulk of our work is found in Sec.~\ref{sec:groverstats}, which covers the probability with which ensemble implementations succeed and uses these as a basis for comparisons with classical probabilistic searches. Finally the appendices contain elaborations of the mathematics behind our results. 


\section{Grover's Search Algorithm: Conventional Implementation}
\label{sec:groversingle}

The simplest unstructured database search problem involves a database that conceals a single marked item in one of $N$ possible locations, conveniently represented by the integers $\{0,1,\ldots N-1\}$. The task is to find the location of the marked item, $s,$ with the help of an oracle, 
\begin{equation}
	f(x) :=
 \begin{cases}
  0 & \text{if $x \neq s$} \\
  1 & \text{if $x = s$}.
 \end{cases}
	\label{eq:oracle}
\end{equation} 
A classical sequential search proceeds by evaluating the oracle at distinct database locations and on average requires $N/2$ such oracle queries to determine the marked item's location. 

The Grover search algorithm~\cite{grover97,nielsen00} requires $n=\lceil \log_2{N}\rceil$ qubits whose computational basis states $\{ \ket{x},  x=0,\ldots 2^n -1 \}$ can be regarded as representing database locations. The oracle is implemented via a unitary transformation, defined on the computational basis states as
\begin{equation}
 \hat{U}_f \ket{x} := \left( -1 \right)^{f(x)} \ket{x}
\end{equation}
and extended linearly to all superpositions. The oracle is easily extended to the database locations $\{N,N+1,\ldots, 2^n-1 \}$ by setting $f(x)=0$ at these points. For this reason we shall restrict discussion of the quantum algorithm to cases where $N=2^n$ for some integer $n$. The algorithm unfolds by initializing the qubits to the state
\begin{equation}
 \ket{\psi_i} = \frac{1}{\sqrt{2^n}}\sum_{x=0}^{N-1} \ket{x}
\end{equation} 
followed by repeated applications of the \emph{Grover iterate,} $\hat{G} := \hat{D} \hat{U}_f,$ where the unitary ``inversion about the average'' operation is  
\begin{equation}
 \hat{D} \left( \sum_{x=0}^{N-1} c_x \ket{x} \right) 
   := \sum_{x=0}^{N-1} \left( -c_x + 2 \left< c \right> \right) \ket{x}
\end{equation}
with $\left< c \right> =\sum_{x=0}^{N-1} c_x / N$. Each application of $\hat{G}$ includes one oracle query. For a database containing only one marked item, to which case we restrict our consideration, a standard analysis~\cite{nielsen00,zalka99} demonstrates that after $q$ applications of $\hat{G}$ the state of the system is
\begin{equation}
  \ket{\psi} = \alpha_q \, \ket{s} + 
               \beta_q \, \frac{1}{\sqrt{N-1}} \sum_{x \neq s} \ket{x}
  \label{eq:stdfinal}
\end{equation}
where 
\begin{align}
	\alpha_q & := \sin{\left( \frac{2 q + 1}{2}\, \theta \right)}, \label{eq:alpha} \\
	\beta_q & := \cos{\left( \frac{2 q + 1}{2}\, \theta \right)}
\end{align}
and $0 \leqslant \theta \leqslant \pi$ depends on $N$ via
\begin{equation}
    \theta = \arccos{\left(1 - \frac{2}{N}\right)}.	
    \label{eq:theta}
\end{equation}
The probability that the final computational basis measurement will yield $s$ is
\begin{equation}
 \Pr{\left( \textrm{correct} \right)} = \left| \innerprod{s}{\psi} \right|^2 
                                      = \sin^2{\left( \frac{2 q + 1}{2}\, \theta \right)}
\end{equation}
and the measurement returns the marked item's location with certainty whenever $(2 q + 1)\, \theta /2$ is an odd multiple of $\pi/2.$ The lowest value of $q$ which ensures this is $\pi/(2\theta) -1/2$ but for an arbitrary database size, $\theta$ may not be such that this is integral. The standard protocol for the Grover algorithm requires that the number of applications of $\hat{G}$ is
\begin{equation}
	q_\mathrm{std} = \nint{\left(\frac{\pi}{2 \theta} - \frac{1}{2}\right)}
\end{equation}
where $\nint{(x)}$ is the nearest integer to $x.$ The probability with which a computational basis measurement will yield $s$ correctly can be ascertained by noting that $\pi/(2\theta) -1 \leqslant q_\mathrm{std} \leqslant \pi/(2\theta).$ At either extreme, the probability that the computational basis measurement yields $s$ is $\cos^2{(\theta/2)}= 1-1/N;$ thus the probability of successfully locating the marked item after $q_\mathrm{std}$ oracle invocations is at least $1-1/N.$ It can be shown that, throughout this range, the probability that the final computational basis measurement yields $s$ is bounded from below by the probability at the extremes of the range.  Typically, the search problem is considered for $N \gg 1,$ in which case the Grover algorithm yields the marked item's location with near certainty. The number of oracle invocations required for the standard algorithm can be approximated from Eq.~\eqref{eq:theta} 
\begin{equation}
    \theta \approx \frac{2}{\sqrt{N}} \left[ 1 + \order{\left(\frac{1}{N}\right)} \right]	
           \approx \frac{2}{\sqrt{N}}
\end{equation}
and thus
\begin{equation}
	q_\mathrm{std} = \nint{\left(\frac{\pi}{4}\, \sqrt{N} - \frac{1}{2}\right)} 
	               \approx \nint{\left(\frac{\pi}{4}\, \sqrt{N} \right)}.
\end{equation}


\section{Grover search algorithm on a ensemble of quantum systems}
\label{sec:groverensemble}

We consider an ensemble implementation of Grover's algorithm starting with a pseudopure state of the type given by Eq.~\eqref{eq:pseudopure}. Eqs.~\eqref{eq:pseudopurefinal} and~\eqref{eq:stdfinal} imply that the density operator after $q$ applications of $\hat{G}$ and prior to measurement is
\begin{widetext}
\begin{align}
	\hat{\rho}_\mathrm{final} & = \frac{\left(1-\eps\right)}{2^{n}}\, \hat{I}^{\otimes n}
	                            + \eps 
	                              \lvert\alpha_q \rvert^2\,
	                              \ket{s} \bra{s}
	                            + \eps
	                              \frac{\alpha_q\beta_q^*}{\sqrt{N-1}}\,
	                              \sum_{x \neq s}
	                              \ket{s} \bra{x}
	                            + \eps
	                              \frac{\alpha_q^*\beta_q}{\sqrt{N-1}}\,
	                              \sum_{x \neq s}
	                              \ket{x} \bra{s}
	                            + \eps
	                            \frac{\lvert \beta_q \rvert^2}{N-1}
	                              \sum_{x \neq s}
	                              \sum_{y \neq s}
	                              \ket{x} \bra{y}.
	                           \label{eq:groverfinal}
\end{align}
%
The standard protocol for determining the algorithm outcome, and one which eliminates the term proportional to $\hat{I},$ is based on the \emph{expectation values of traceless single qubit observables}~\cite{knill98,chuang98a}. For the $j^\mathrm{th}$ qubit, single qubit expectation values can be computed from  $\hat{\rho}_\mathrm{final\, red}^{(j)},$ the final reduced density operator for the qubit. For Grover's algorithm, Eq.~\eqref{eq:groverfinal} implies that, after $q$ applications of $\hat{G},$
%
\begin{align}
	\hat{\rho}_\mathrm{final\, red}^{(j)} = & \frac{1}{2}\, \left[ 1 - \eps\, 
	                                                  \frac{\lvert \alpha_q \rvert^2\, N -1 }{N-1} \right]\, 
	                                                  \hat{I} 
	                                          + \eps\, \frac{\lvert \alpha_q \rvert^2\, N -1 }{N-1}\,
	                                                 \ket{s_j}\bra{s_j} 
	                                       \nonumber \\
	                                        & + \eps
	                                            \left( \frac{N}{2} -1 \right)\, 
	                                            \frac{1- \lvert \alpha_q \rvert^2}{N-1}\,
	                                            \biggl[ \ket{s_j}\bra{s_j \oplus 1} 
	                                                 + \ket{s_j \oplus 1}\bra{s_j} \biggr]
	                                       \nonumber \\
	                                        & + \eps 
	                                            \frac{\alpha_q^* \beta_q}{\sqrt{N-1}}\,
	                                            \ket{s_j \oplus 1}\bra{s_j}
	                                          + \eps 
	                                            \frac{\alpha_q \beta_q^*}{\sqrt{N-1}}\,
	                                            \ket{s_j}\bra{s_j \oplus 1}
	                                            \label{eq:reddensityop}
\end{align}
\end{widetext}
where $s_j$ is the $j^\mathrm{th}$ bit of $s.$ The expectation value,
\begin{align}
 \expv{\sigma_z}^{(j)}  & = \Trace{\left( \hat{\sigma}_z \hat{\rho}_\mathrm{final\, red}^{(j)} \right)}	
	 																				\nonumber \\
	 									    & = \left( -1 \right)^{s_j}\, \eps\, \frac{\lvert \alpha_q \rvert^2\, N -1 }{N-1}
 \label{eq:expvaloutcome}
\end{align}
depends on $s_j.$ Thus $s_j$ can be determined whenever the \emph{expectation value of $\hat{\sigma}_z$ can be measured for the $j^\mathrm{th}$ qubit} and the measurement resolution allows for distinction between $+\eps\,(\lvert \alpha_q \rvert^2\, N -1)/(N-1)$ and $-\eps\,(\lvert \alpha_q \rvert^2\, N -1)/(N-1).$ This opens the possibility for a truncated version of Grover's algorithm, which uses fewer oracle invocations (i.e.\ $q < q_\mathrm{std}$) than required by the standard version of Grover's algorithm, subject to the proviso that $\lvert \alpha_q \rvert$ is large enough to distinguish between the two possible expectation values~\cite{collins02}. If these requirements are satisfied, then an ensemble implementation based on expectation values will be \emph{deterministic} in the sense that a given value of $s$ will always yield the same outcomes and the value of $s$ can be determined with certainty.

For an ensemble containing a finite number of members, $M,$ precise expectation value measurements are idealizations. Their purpose is to suggest protocols that produce algorithm outputs based on appropriate sample averages of projective measurement outcomes for individual ensemble members. The key notion is that, for a sufficiently large $M$, sample averages of projective measurement outcomes give close approximations to expectation values. The typical protocol has already been described for single bit output algorithms~\cite{anderson05} and is readily extended to multiple bit output algorithms. First, a computational basis measurement is performed across all qubits for each ensemble member. Consider the $j\mathrm{th}$ qubit of ensemble member $k$ (where $1 \leqslant k \leqslant M$). The projectors for the two possible computational basis measurement outcomes are $\hat{P}_{s_j} := \ket{s_j}\bra{s_j}$ and $\hat{P}_{\overline{s}_j} := \ket{s_j \oplus 1}\bra{s_j \oplus 1}$ and the corresponding measurement outcome are scaled to $z_k^{(j)}=+(-1)^{s_j}$ and $z_k^{(j)}=-(-1)^{s_j}$ respectively; this translates between the computational basis measurement outcomes  and the eigenvalues of $\sigma_z.$ Second, the sample average for the $j\mathrm{th}$ qubit is computed
\begin{equation}
 \bar{z}^{(j)}:=\frac{1}{M}\sum_{k=1}^{M} z_k^{(j)}
 \label{eq:sampave}
\end{equation}
and this approximates $\expv{\sigma_z}^{(j)}.$ Finally, if $\bar{z}^{(j)}>0$ (or conversely $\bar{z}^{(j)} < 0$) then this suggests that $\expv{\sigma_z}^{(j)} > 0,$ (or $\expv{\sigma_z}^{(j)} < 0$) which, according to Eq.~\eqref{eq:expvaloutcome} implies $s_j=0$ (or $s_j=1$).  Thus a \emph{candidate, $c:=c_n \ldots c_1$, for the marked item's location} is assigned via: 
\begin{equation}
  \renewcommand{\arraystretch}{1.5}
  \begin{array}{ll}
     \bar{z}^{(j)} > 0 \; \Rightarrow \;  & c_j = 0\\
     \bar{z}^{(j)} = 0 \; \Rightarrow \;  & \left\{ 
                                                \textrm{\parbox{2.25in}{$c_j=0$ with probability $1/2$,\\ 
                                                                        $c_j=1$ with probability $1/2$ and}}
                                            \right.\\
     \bar{z}^{(j)} < 0 \; \Rightarrow \;  & c_j =1.\\
  \end{array}
\label{eq:sampprotocol}
\end{equation}
The protocol of Eqs.~\eqref{eq:sampave} and~\eqref{eq:sampprotocol} is equivalent to taking a bitwise majority vote of the computational basis measurement outcomes over the entire ensemble. If, for a given bit, more ensemble members return $+1$ than $-1$ after computational basis measurements, then the intuition is that $s_j=0$ since the outcome $(-1)^{s_j}$ is more likely than $-(-1)^{s_j}.$ The string $c:=c_n \ldots c_1$ constructed in this fashion constitutes the ensemble algorithm's output.

The probabilities with which single qubit outcomes occur are readily determined from the reduced density operator of Eq.~\eqref{eq:reddensityop}. Specifically the probability that an ensemble member yields the ``correct'' value is
\begin{align}
	\Pr{\left(z_k^{(j)} = +(-1)^{s_j}\right)} & = \Trace{\left( \hat{\rho}_\mathrm{final\, red}^{(j)} \hat{P}_{s_j} \right)} 
	                                 \nonumber \\
	                             & = \frac{1+\epseff}{2}
	\label{eq:probcorrect}  
\end{align}
where the \emph{effective polarization} is
\begin{equation}
	\epseff := \eps\, \frac{\lvert \alpha_q \rvert^2\, N -1 }{N-1}.
	\label{eq:epseffective}
\end{equation}
Similarly  the probability that an ensemble member yields the ``incorrect'' value is
\begin{equation}
	\Pr{\left(z_k^{(j)} = -(-1)^{s_j}\right)}  = \frac{1- \epseff}{2}. 
	\label{eq:probincorrect}
\end{equation}
Unless both $\eps=1$ and $\lvert \alpha_q \rvert^2\ = 1,$ the protocol of Eqs.~\eqref{eq:sampave} and ~\eqref{eq:sampprotocol} can give  $c_j \neq s_j,$ thus causing the algorithm to fail to locate the marked item correctly. In this sense the algorithm is \emph{probabilistic.} Even for perfect polarization, this is generally unavoidable since, for the standard number of invocations of the Grover iterate, i.e.\ $q=q_\mathrm{std},$ $\lvert \alpha_q \rvert^2\, \neq 1.$ However, for the standard number of iterations,
\begin{equation}
  1-\frac{1}{N} \leqslant \lvert \alpha_q \rvert^2\, \leqslant 1
\end{equation}
giving
\begin{equation}
	\eps \left( 1-\frac{1}{N-1} \right) \leqslant  \epseff \leqslant \eps
\end{equation}
and as $N \rightarrow \infty,$ $\epseff \approx \eps.$ Thus for the \emph{typical scenario,} for which $N \gg 1$ and $q=q_\mathrm{std}$ in ensemble implementations the probabilistic aspects will be determined predominantly by the polarization. We shall henceforth only consider the typical scenario, taking $\epseff = \eps$ and $q = \pi \sqrt{N}/4$.

The protocol of Eq.~\eqref{eq:sampprotocol} is one in which it is assumed that it is possible to distinguish a discrepancy as small as $1$ between the numbers of projective measurement outcomes returning $+1$ and $-1$ respectively. This is the \emph{best resolution case} and one can generalize the situation to consider cases where the resolution is worse as has been done for single-bit output algorithms~\cite{anderson05}. In this article we shall restrict the discussion to the best resolution case.


\section{Statistical performance of the ensemble quantum search algorithm}
\label{sec:groverstats}

The probabilistic nature of quantum algorithms on ensembles motivates performance comparisons, based on the probability with which each algorithm fails or succeeds to correctly identify the marked item's location, with probabilistic classical competitors. Of the possible scenarios, we shall consider the probability of correctly identifying \emph{every bit} $s_1, \ldots, s_n$ of the marked item's location. Eqs~\eqref{eq:probcorrect} and~\eqref{eq:epseffective} suggest that this will depend on the database size $N$ and the polarization $\eps$ (for the typical scenario $q$ is determined by $N$). Also, as with any sample average, it is to be expected that the failure or success probability will depend on the ensemble size $M$. Denote the probability with which the ensemble quantum algorithm correctly identifies \emph{every bit of $s$} by $\psallq{\eps}{M}{N}$ and the probability with which it fails (i.e.\ at least one bit of $s$ is incorrectly identified) by $\pfallq{\eps}{M}{N} = 1 - \psallq{\eps}{M}{N}.$ Methods for calculating and approximating these have been provided for ensemble quantum algorithms with single bit outputs~\cite{anderson05}. The key difference here is that Grover's algorithm's output involves multiple qubits and that measurement outcomes for \emph{distinct qubits in an individual ensemble member are correlated}. Thus, although we can still use the methods for single bit outputs~\cite{anderson05} to determine the probability with which any single bit value will be correctly identified, the fact that the measurement outcomes for different qubits are not statistically independent means that success probabilities for multiple output bit cases will not necessarily be simple combinations of those for the single output bit case. 

To illustrate this point, consider the case where $n=2$ (i.e.\ $N=4$) and the marked item is located at $s=11$ (in binary notation). Eq.~\eqref{eq:theta} implies that $\theta = \pi/3$ giving $q_\mathrm{std} = 1.$ Then Eq.~\eqref{eq:alpha} implies $\ket{\psi_f} = \ket{s} = \ket{11}.$ For an ensemble realization, Eq.~\eqref{eq:epseffective} gives $\epseff = \eps.$ This yields the joint probability distribution for various measurement outcomes on both qubits of a single ensemble member of Table~\ref{tab:jointprob}.
\begin{table}[b]
	\centering
	    \renewcommand{\arraystretch}{2.5}
	    \begin{tabular}{||c|c|c||}
	      \hline \hline
		  Bit 2 Outcome & Bit 1 Outcome & Probability \\
		  \hline
	      0 & 0 & $\dfrac{1-\eps}{4}$ \\
		  \hline
	      0 & 1 & $\dfrac{1-\eps}{4}$ \\
		  \hline
	      1 & 0 & $\dfrac{1-\eps}{4}$ \\
		  \hline
	      1 & 1 & $\dfrac{1+ 3\eps}{4}$ \\
	      \hline \hline
		\end{tabular}
	\caption{Joint probability distribution for $N=4$ and $s=11$ (binary).}
	\label{tab:jointprob}
\end{table}

The marginal probability distributions for single qubit outcomes are identical; for the $k^\mathrm{th}$ ensemble member and the $j^\mathrm{th}$ bit, $\Pr{( z_k^{(j)} = 0) } = (1-\eps)/2$ and $\Pr{(z_k^{(j)} = 1) } = (1+\eps)/2.$ Thus the probability distributions for the individual qubit outcomes are \emph{not} independent except in the trivial cases of $\eps=0$ or $\eps=1.$ Let $k_1$ denote the number of times that the outcome for qubit~1 is 1 and $k_2$ the number of times that it is 1 for qubit 2. 
When the number of ensemble members is odd the ensemble algorithm succeeds provided that $k_1>M/2$ \emph{and}  $k_2> M/2.$ It is shown in appendix~\ref{app:twoqubitssuccess} that the probability with which single qubit measurements yield $k_1$ and $k_2$ is
\begin{widetext}
\begin{equation}
 \Pr{(k_2,k_1)} = \frac{1}{4^M}\, \sum_{l=\max{(0,k_1+k_2-M)}}^{\min{(k_1,k_2)}} 
                  (M-k_1-k_2+l,k_1-l,k_2-l,l)!\, 
                  \left( 1-\eps \right)^{M-l}\,
                  \left( 1+3\eps \right)^{l}
\end{equation}
\end{widetext}
where $(n_1,n_2,n_3,n_4)! = (n_1+n_2+n_3+n_4)!/(n_1!n_2!n_3!n_4!).$ Then
\begin{equation}
  \psallq{\eps}{M}{N} = \sum_{k_1=\frac{M+1}{2}}^{M}
                           \sum_{k_2=\frac{M+1}{2}}^{M} \Pr{(k_2,k_1)}.
  \label{eq:twoqubitssuccessall}
\end{equation}
For algorithms with single bit outputs, the expression analogous to Eq.~\eqref{eq:twoqubitssuccessall} is a cumulative binomial distribution, which can be expressed as an incomplete beta function, greatly simplifying various calculations and inferences~\cite{anderson05}. We are unaware of comparable techniques for re-expressing Eq.~\eqref{eq:twoqubitssuccessall} in cases involving more than one output bit. 

Nevertheless certain general conclusions are possible. If $\eps=0,$ the ensemble is in the maximally mixed state throughout the algorithm. Thus for a given qubit, a computational basis measurement yields either outcome with equal probability. The procedure for deciding the algorithm output amounts to unbiased guessing on each bit and these outcomes are no longer correlated. It follows that 
\begin{equation}
 \psallq{0}{M}{N} = \frac{1}{2^n} = \frac{1}{N}.
 \label{eq:psall0}
\end{equation} 
At the other extreme $\eps =1$ and $\rho_\mathrm{final} = \ket{\psi_\mathrm{final}}\bra{\psi_\mathrm{final}}.$ The fact that in the typical scenario $\lvert \alpha_{q_\mathrm{std}} \rvert^2$ is not necessarily $1$ implies that in general, $\epseff \neq 1,$ and the measurement outcomes are correlated. However, $\epseff \approx 1$ and 
\begin{equation}
 \psallq{1}{M}{N} \approx 1.
 \label{eq:psall1}
\end{equation} 
for $N \gg 1.$ A more precise lower bound on $\psallq{1}{M}{N}$ can be obtained via Eqs.~\eqref{eq:pfone1bound} and~\eqref{eq:pfailbounds}; this implies the result of Eq.~\eqref{eq:psall1} for $N \gg 1.$

The absence of techniques for computing or approximating $\psallq{\eps}{M}{N}$ for more than a single output bit motivates the search for bounds on this quantity or, equivalently and more conveniently, $\pfallq{\eps}{M}{N}.$ The strategy is to bound this in terms of the \emph{single bit failure probability}, i.e.\ the marginal probability with which the algorithm fails to identify a given single qubit correctly regardless of the output for the other bits. Assuming, without loss of generality that the marked item is located at $s= 11\ldots 11,$ Eq.~\eqref{eq:pseudopurefinal} implies that the joint probability distribution is symmetrical under interchanges of qubits. Thus the single bit failure probability is identical for all qubits and will be denoted $\pfoneq{\eps}{M}{N}$. The bounding relationship, central to the rest of this article,  between the single bit failure probability and the probability with which the algorithm fails is 
\begin{equation}
 p_\mathrm{fail\; one}^{\phantom{fail\! one} \mathrm{q}} 
  \leqslant 
    p_\mathrm{fail\; all}^{\phantom{fail\! all} \mathrm{q}} 
  \leqslant 
 n\, p_\mathrm{fail\; one}^{\phantom{fail\! one} \mathrm{q}}
 \label{eq:pfailbounds}
\end{equation}
where the arguments, $(\eps,M,N),$ of the probabilities  are identical and have been omitted for convenience (see appendix~\ref{app:failurebounds} for a proof).

Following methods established for single bit output algorithms~\cite{anderson05}, 
%
\begin{align}
	\pfoneq{\eps}{M}{N} & = 
	                        \nonumber \\
	                    & \hspace{-10ex} \phantom{+}  \frac{1}{2}\, \cbd{M}{\Mmin}{\effmn}{\effpn} 
	                        \nonumber \\
	                    & \hspace{-12ex} + \frac{1}{2} \cbd{M}{M- \Mmin+1}{\effmn}{\effpn}
\label{eq:pfailqgen}
\end{align}
%
where $\epseff$ is given by Eq.~\eqref{eq:epseffective} and
\begin{equation}
	\Mmin := \left\lceil \frac{M+1}{2} \right\rceil.
\end{equation}
The single bit failure probability is a monotonically decreasing function of $\eps$~\cite{anderson05} with $\pfoneq{0}{M}{N}= 1/2$ and 
\begin{equation}
	\pfoneq{1}{M}{N} \leqslant \sqrt{\frac{2}{\pi M}}\, 
	                              \frac{2(N-1)}{2N-3}\,
	                              \left( \frac{2}{N-1} \right)^{M/2}
	\label{eq:pfone1bound}
\end{equation}
whenever $M \gg 1$ (see appendix~\ref{app:epsone}). Thus for $N \gg 1,$ $\pfoneq{1}{M}{N} \approx 0.$ For fixed polarization $\eps$ and odd $M,$ $\pfoneq{\eps}{M}{N} = \pfoneq{\eps}{M+1}{N}$ and $\pfoneq{\eps}{M}{N} \geqslant \pfoneq{\eps}{M+2}{N}$ with equality only if $\epseff=0,1$~\cite{anderson05}.

The exact expression of Eq.~\eqref{eq:pfailqgen} is not conducive to rapid numerical evaluation and is less useful for assessing the behavior of $\pfoneq{\eps}{M}{N}$ than certain alternatives. It can easily be verified by repeated integration by parts that 
\begin{equation}
	\sum_{k=m}^n \binom{n}{k} p^k (1-p)^{n-k} = I_p(m,n-m+1)
\end{equation}
where
\begin{equation}
	I_p(x,y) := \frac{\Gamma(x+y)}{\Gamma(x) \Gamma(y)}\, 
	            \int_0^p t^{x-1}(1-t)^{y-1}
	            \label{eq:beta}
\end{equation}
is the incomplete beta function. Applied to Eq.~\eqref{eq:pfailqgen} this yields an alternative exact expression, 
%
\begin{equation}
  \begin{split}
	\pfoneq{\eps}{M}{N} = & \frac{1}{2}
	                        \bigl[ I_p(\Mmin,M-\Mmin+1)
	                              \\
	                              & \quad + I_p(M-\Mmin+1,\Mmin) 
	                             \bigr]
	                           \end{split}
	                       \label{eq:pfqbeta}
\end{equation}
%
where $p=(1-\eps)/2.$ For $M \gg 1,$ Eqs.~\eqref{eq:beta} and~\eqref{eq:pfqbeta} can be combined (see appendix~\ref{app:gaussapprox}) to give the approximation
\begin{equation}
	\pfoneq{\eps}{M}{N} \approxeq \frac{1}{2} - \frac{1}{\sqrt{2 \pi}}\, \int_0^{\eps\sqrt{M}} e^{-t^2/2} \mathrm{d}t.
	                              \label{eq:pfqerf}
\end{equation}
This is closely related to the standard error function and is more suitable for numerical computation than the exact expression of Eq.~\eqref{eq:pfailqgen}.

\subsection{Statistical comparison of the ensemble quantum and classical sequential search algorithms}

The ensemble quantum algorithm is probabilistic and it is reasonable to compare it in terms of success probabilities to a probabilistic classical sequential search. A deterministic classical sequential search proceeds by evaluating the oracle, $f(x),$ at distinct database locations $x,$ terminating whenever $f(x)=1.$ We consider a probabilistic sequential search in which the oracle is evaluated at $Q$ distinct randomly chosen database locations. The probability with which this fails to correctly identify the marked item is
\begin{equation}
	 \pfallc{Q}{N} = 1-\frac{Q}{N}
	 \label{eq:pfclassical}
\end{equation}
provided that $Q \leqslant N-2$ since a search with $N-1$ queries will yield the marked item's location with certainty. We shall compare the two types of searches by requiring that each use \emph{the same resources measured in terms of the aggregate number of oracle invocations} (we briefly discuss other possibilities later). Thus we regard an ensemble implementation invoking $q$ oracle invocations, each on $M$ ensemble members, as using equivalent resources to a classical sequential with $Q = qM$ oracle queries. The central issue in this article is to compare the two types of searches provided that $Q=qM$ and to determine conditions under which the ensemble implementation of the quantum algorithm outperforms the probabilistic classical sequential search, i.e.\ such that 
\begin{equation}
	\pfallq{\eps}{M}{N} < \pfallc{qM}{N}.
\end{equation}
The \emph{critical polarization} $\epscrit,$ which demarcates the boundary between the two types of algorithm, is defined implicitly via
\begin{equation}
 \pfallq{\epscrit}{M}{N} = \pfallc{qM}{N}.
 \label{eq:epscritdef}
\end{equation}
 In general $\epscrit$ can be expected to depend on $M$ and $N$ and will be denoted $\epscrit(M,N).$  Eqs.~\eqref{eq:psall0},~\eqref{eq:psall1} and ~\eqref{eq:epscritdef} imply that a critical polarization in the range $0 \leqslant \epscrit \leqslant 1$ can only exist provided that 
\begin{equation}
	1 \leqslant qM \leqslant N.
\end{equation}
The left bound is trivially satisfied while the right bound implies an upper limit on the ensemble size,
\begin{equation}
	\Mmax = \left\lfloor \frac{N}{q} \right\rfloor.
	\label{eq:Mrange} 
\end{equation}
For $M > \Mmax,$ $\pfallc{qM}{N} = 0$ and the classical algorithm always outperforms the quantum algorithm, except in some (but not all) situations for which $\eps=1.$ For $M < \Mmax$ the critical polarization exists since $\pfallq{\eps}{M}{N}$ is a continuous function of $\eps$ and the classical failure probability lies within the bounds for the quantum failure probability. 

The difficulties in calculating $\psallq{\eps}{M}{N}$ are equivalent to those in calculating $\pfallq{\eps}{M}{N}$ and we resort to calculating bounds, originating from the inequality of Eq.~\eqref{eq:pfailbounds}, for the critical polarization. We shall show that 
\begin{equation}
   \epscritone(M,N) \leqslant  \epscrit(M,N) \leqslant \epscritall(M,N)
   \label{eq:epscritbound}
\end{equation}
where the \emph{necessary critical polarization,} $\epscritone,$ satisfies
\begin{equation}
	\pfoneq{\epscritone}{M}{N} = \pfallc{qM}{N}.
	\label{eq:epscritonedef}
\end{equation}
and the \emph{sufficient critical polarization}, $\epscritall,$ satisfies
\begin{equation}
	\pfoneq{\epscritall}{M}{N} = \frac{1}{n}\, \pfallc{qM}{N}.
\end{equation}
Each of these depend on $M$ and $N$ and they will sometimes be denoted $\epscritone(M,N)$ and $\epscritall(M,N).$ As the necessary and sufficient critical polarizations are defined via the single bit failure probability, existing techniques~\cite{anderson05} can be used to demonstrate their properties and calculate them.  

The existence of the necessary critical polarization follows from the fact~\cite{anderson05} that $\pfoneq{\eps}{M}{N}$ is a monotonically decreasing function of $\eps$ ranging from $0$ to $1/2.$ Thus a unique value of the necessary critical polarization exists whenever $\lceil N/2q \rceil \leqslant M \leqslant \Mmax.$ Note that in the typical scenario this implies $ \Mmax/2 \leqslant M \leqslant \Mmax.$ For $M \leqslant \lceil  N/2q \rceil,$ the classical sequential search fails with probability larger than $1/2$ and thus the ensemble quantum algorithm always outperforms it, giving $\epscritone(M,N) =0.$ Similarly, the sufficient critical polarization exists provided that $ (1-n/2)N/q \leqslant M \leqslant \Mmax.$ The left hand inequality is satisfied whenever $n\geqslant 2$ as it does in the typical scenario. Here a unique sufficient polarization exists when $ 0 \leqslant M \leqslant \Mmax.$

The term ``necessary'' stems from considering $\eps < \epscritone,$ in which case the monotonic behavior of $\pfoneq{\eps}{M}{N}$ with respect to $\eps$ and Eqs~\eqref{eq:pfailbounds} and~\eqref{eq:epscritonedef} imply 
\begin{equation}
	  \pfallq{\eps}{M}{N} > \pfallc{qM}{N}; 
\end{equation}
here the classical algorithm succeeds with greater probability than the quantum ensemble algorithm. In this sense, $\epscritone$ establishes a lower bound on $\epscrit.$ 

The term ``sufficient'' arises by considering,  $\eps > \epscritall,$ for which a similar argument implies 
\begin{equation}
	\pfallq{\eps}{M}{N} < \pfallc{qM}{N};  
\end{equation}
here the quantum ensemble algorithm succeeds with greater probability than the classical sequential search using the same resources. In this regard $\epscritall$ is an upper bound on $\epscrit.$ This establishes Eq.~\eqref{eq:epscritbound}.

\begin{figure}[h]
  \includegraphics[scale=1]{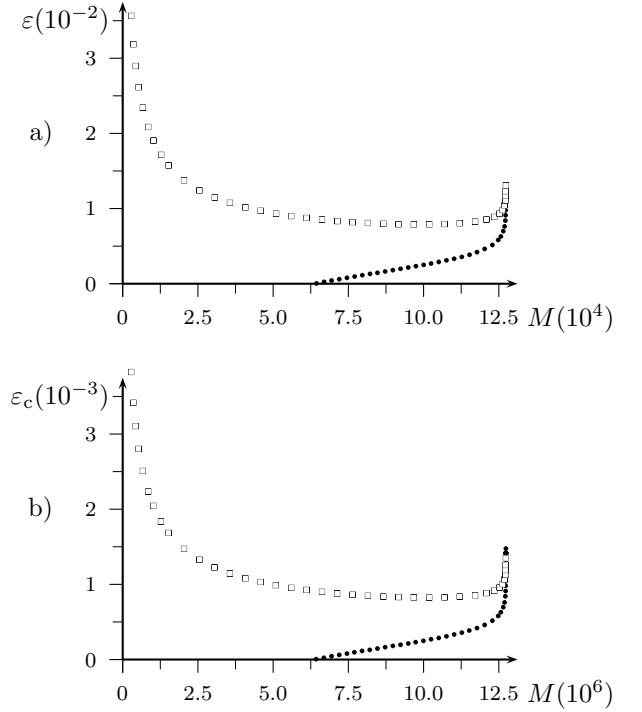}
  \caption{Critical polarization for a) $N=10^{10}$ and b) $N=10^{14}.$ Open squares indicate sufficient polarization and solid dots necessary polarization. Note the distinct horizontal and vertical scales. The rightmost point in each plot corresponds to $M = \Mmax$ for the relevant value of $N$. The critical polarization, $\epscrit,$ lies between the two curves.}
    \label{fig:pol}
 \end{figure}

The task now becomes one of computing the two bounding critical polarizations. Eqs.~\eqref{eq:pfclassical} and~\eqref{eq:epscritonedef} with $Q=qM$ imply that
\begin{align}
	\pfoneq{\epscritone}{M}{N} & =  1-\frac{qM}{N} 
	                                \nonumber \\
	                           & \approx 1-\frac{\pi M }{4\sqrt{N}}.
	\label{eq:necpol}
\end{align}
For fixed $N,$ Eq.~\eqref{eq:necpol} can be solved numerically for $\epscritone(M,N).$ Data, computed via Eq.~\eqref{eq:pfqbeta} using MATHEMATICA's implementation of the incomplete beta function, is plotted in Fig.~\ref{fig:pol}. 

In general it can be shown (see appendix~\ref{app:necvsM}) that $\epscritone(M,N)$ is a monotonically increasing function of $M$ with $\epscritone \rightarrow 1$ as $M \rightarrow \Mmax.$ A more important consideration is the scaling behavior of $\epscritone(M,N)$ with respect to $N$ while fixing $M$ suitably. The similarity in the data generated for the two values of $N$ illustrated in Fig.~\ref{fig:pol} suggests an approach based on $\epscritone$ as a function $M/\Mmax$ for different values of $N.$ In fact, it is easily shown that 
\begin{equation}
  \pfallc{qM}{N} \approx 1 - \frac{M}{\Mmax}
\end{equation}
where $\Mmax$ depends on $q$ and $N$ via Eq.~\eqref{eq:Mrange}. Thus the necessary critical polarization is approximately determined via
\begin{equation}
  \pfoneq{\epscritone}{M}{N} = 1 - \frac{M}{\Mmax}.
\end{equation}
For $N \gg 1,$ Eqs.~\eqref{eq:epseffective} and~\eqref{eq:pfailqgen} imply that the left hand side is independent of $N$ and $q$ and these only enter into the problem via $\Mmax.$ Our strategy for determining the scaling behavior of $\epscritone$ with respect to $N$  is to consider fixed $M/\Mmax$ as $N$ varies. This is equivalent to maintaining an ensemble size needed to sustain a fixed failure probability for the classical algorithm. In the typical scenario $\Mmax \approx \pi \sqrt{N}/4$ and upon multiplying $N$ by a factor $\gamma,$ the ratio $M/\Mmax$ can only be held constant by multiplying $M$ by a factor of $\sqrt{\gamma}.$ More explicitly, for any $\gamma>0,$ Eq.~\eqref{eq:necpol} implies that
\begin{equation}
  \begin{split}
	 \mathrm{} & \pfoneq{\epscritone(\sqrt{\gamma}M,\gamma N)}{\sqrt{\gamma}M}{\gamma N} 
	             \nonumber \\
	           & = \pfoneq{\epscritone(M,N)}{M}{N}
	\end{split}
\end{equation}
and the approximation of Eq.~\eqref{eq:pfqerf} implies 
\begin{equation}
	\epscritone(\sqrt{\gamma}M,\gamma N) \approx \gamma^{-1/4} \epscritone(M,N)
	\label{eq:epsonescaling}
\end{equation}
or equivalently
\begin{equation}
	\epscritone(M,\gamma N) \approx \gamma^{-1/4} \epscritone(M/\sqrt{\gamma},N).
	\label{eq:epsonescalingtwo}
\end{equation}
This relationship is evident in the numerical data; the plots of Fig.~\ref{fig:pol} illustrate an instance where $\gamma=10^4.$ 

The sufficient polarization satisfies
\begin{align}
	\pfoneq{\epscritall}{M}{N} & = \frac{1}{n}\, \left( 1-\frac{qM}{N} \right)
			                           \nonumber \\
			                       & \approx \frac{1}{n}\, \left( 1-\frac{M }{\Mmax} \right)
	\label{eq:suffpol}
\end{align}
Data generated via numerical solution of Eq.~\eqref{eq:suffpol}, using the incomplete beta function representation of Eq.~\eqref{eq:pfqbeta}, is illustrated in Fig.~\ref{fig:pol}. A notable feature of the data is that for much of the region in which $\Mmax/2 \leqslant M \leqslant \Mmax,$ the necessary and sufficient critical polarizations are within an order of magnitude of each other; if all that is of interest is the typical order of magnitude of $\epscrit$ for a given $N$ then it suffices to compute either $\epscritone$ or $\epscritall$ for $M$ a moderately large fraction of $\Mmax.$ By this argument one can attain an order of magnitude estimate of the critical polarization, which lies between the bounding critical polarizations. Additionally, for much of the range $0 \leqslant M \leqslant \Mmax,$ the numerical data indicates that the sufficient polarization is roughly constant.

In the limiting case $\epscritall \rightarrow 1$ as $M \rightarrow \Mmax.$ Also as $M \rightarrow 1,$ $\pfoneq{\epscritall}{M}{N} \rightarrow (1-\eps)/2$ and thus Eq.~\eqref{eq:suffpol} implies that
\begin{equation}
 	\epscritall \rightarrow 1 - \frac{2}{n}\, \left( 1- \frac{\pi}{4 \sqrt{N} } \right) \approx 1
\end{equation}
in the typical scenario. The scaling of the sufficient critical polarization is more complicated than that of the necessary critical polarization as a result of the additional factor of $1/n$ in Eq.~\eqref{eq:suffpol}. A similar line of reasoning to that for the necessary polarization gives
%
%
\begin{align}
  & \pfoneq{\epscritall(\sqrt{\gamma}M,\gamma N)}{\sqrt{\gamma}M}{\gamma N} 
    \nonumber \\
  & \quad = \frac{ \left\lceil \log_2{N} \right\rceil}{\left\lceil \log_2{(\gamma N)} \right\rceil \, } \,
	  \pfoneq{\epscritall(M,N)}{M}{N}
 	  \label{eq:critallrations}
\end{align}
%
Here the terms involving logarithms prohibit immediate use of Eq.~\eqref{eq:pfqerf} as was done for the necessary polarizations. However, when $\left\lceil \log_2{\gamma} \right\rceil  \ll \left\lceil \log_2{N} \right\rceil,$ 
\begin{equation}
	\epscritall(\sqrt{\gamma}M,\gamma N) \approx \gamma^{-1/4} \epscritall(M,N),
	\label{eq:epsallscaling}
\end{equation}
since the factor involving logarithms on the right of Eq.~\eqref{eq:critallrations} is approximately unity. Again this evident for the case of $\gamma = 10^{4}$ as illustrated in Fig.~\ref{fig:pol}.

\begin{figure}[h]
  \includegraphics[scale=1]{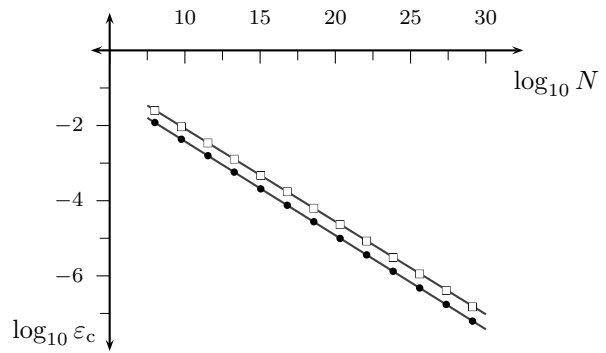}
  \caption{Critical polarization for success probability of $0.90$ Open squares indicate sufficient critical polarization and solid dots necessary critical polarization. The straight lines are obtained by least squares fits. Similar plots are obtained for other values of $p$, the primary difference being that the lines are shifted vertically.}
  \label{fig:givenpol}
 \end{figure}

An additional illustration of the scaling properties of the necessary and sufficient critical polarizations can be attained by computing these at values of $M$ for which the classical algorithm returns a fixed success probability rate as $N$ varies. In order to succeed with probability, $0 \leqslant p \leqslant 1,$ the classical algorithm requires $Q = \left\lceil pN\right\rceil$ oracle queries. The ensemble quantum algorithm uses comparable resources for $M = \left\lceil Q /q\right\rceil \approx 4 p/\pi \sqrt{N} \approx p \Mmax$ ensemble members. Necessary and sufficient critical polarizations can be computed using $N$ and the value of $M$ computed from it as just described. This method of fixing $M$ is the same as that resulting in Eqs.~\eqref{eq:epsonescaling} and~\eqref{eq:epsallscaling}. Dropping the first argument, which is computed from the second and $p$, Eq.~\eqref{eq:epsonescaling} is equivalent to
\begin{equation}
	  \eps_\mathrm{c\; necc}(N) = \left( \frac{N}{N_0} \right)^{-1/4}
	                   \eps_\mathrm{c\; necc}(N_0)
	\label{eq:epsonescalingone}
\end{equation}
where $N_0$ is any fixed database size that is sufficiently large for the approximations to be valid. A similar conclusion applies to the sufficient polarization.
 For $p=0.90,$ numerically generated data is illustrated, using logarithmic scaling, in Fig.~\ref{fig:givenpol}.

Least squares linear fits for these data give
\begin{align}
 \log_{10}{\epscritone} &=0.078 - 0.250 \log_{10}{N} 
                          \nonumber \\
 \log_{10}{\epscritall} &=0.387 - 0.247 \log_{10}{N}
                          \nonumber 
\end{align}
which are consistent with Eqs.~\eqref{eq:epsonescalingone}; the discrepancy in the coefficient of $\log_{10}{N}$ for $\epscritall$ probably arising from the factor of $1/n$ in the definition of the sufficient polarization.

We now briefly consider an alternative method of comparing the two types of algorithm. Suppose that the ensemble algorithm uses fewer resources than the probabilistic sequential search in the sense that $(qM)^\alpha = Q$ where $\alpha >1.$ The fact that the classical succeeds with certainty when $Q \geqslant N-1$ means that, in the typical scenario, the ensemble implementation can only outperform its classical competitor when $\alpha < 2.$ It is easy to show that for the necessary polarization, $0 \leqslant M \leqslant \Mmax,$ where now $\Mmax = \lfloor N^{(1/\alpha)}/q \rfloor$ and for the typical scenario this leaves $\Mmax \approx 4 N^{(1/\alpha -1/2)}/\pi.$ The scaling argument can be repeated by considering multiplication of $N$ by a factor $\gamma >0.$ A constant classical failure probability is then maintained by considering $M$ multiplied by a factor of $\gamma^{(1/\alpha -1/2)}.$ Repeating the argument involving the single bit failure probability for the ensemble quantum algorithm ultimately yields
\begin{equation}
	  \eps_\mathrm{c\; necc}(N) = \left( \frac{N}{N_0} \right)^{[1/4 - 1/(2\alpha)]}
	                   \eps_\mathrm{c\; necc}(N_0)
\end{equation}
with a similar inference for the sufficient polarization.

Current room-temperature, solution state NMR using pseudopure preparation schemes~\cite{chuang98a,jones00,laflamme01} attains typical polarizations of $\eps \sim 10^{-5}.$ The numerical data plotted in Fig.~\ref{fig:pol} and the scaling of Eq.~\eqref{eq:epsonescaling} and~\eqref{eq:epsallscaling} suggest that these standard realizations would only gain an advantage for databases of size $N \gtrsim 10^{22}.$ In this respect, the requirements on the ensemble for an advantageous realization of the Grover search algorithm are less demanding than those for the Deutsch-Jozsa algorithm~\cite{anderson05}, where it was found that $\eps \geqslant 0.866$ was necessary in the best case, regardless of the problem size. It should be noted that the standard number of oracle queries for such database sizes is of the order of $10^{11},$ which forms a lower bound for the number of pulses in sequence of gates; sufficient coherent control over such a number of pulses is evidently beyond current NMR experimental capabilities. Nevertheless, compared to the situation of the ensemble Deutsch-Jozsa algorithm, our results are promising since an increase of one order of magnitude in polarization will result advantages for searches on a database whose size decreases by four orders of magnitude. In fact, for a dramatic increase in polarization such as that produced by newer schemes~\cite{anwar04,anwar04b} producing a large initial polarization of $\eps \sim 0.9$ would be sufficient for modest database sizes where $N \lesssim 10^6.$ Also, to date, all solution state NMR realizations of Grover's algorithm~\cite{chuang98b,jones98a,vdsypen00} have involved databases for which $N \leqslant 8$ and ensemble sizes of $M \sim 10^{20}.$ Thus $M \gg \Mmax$ in these cases and, according to our criteria, the ensemble quantum algorithm uses greater resources than a classical sequential search which will locate the marked item with certainty. 

\section{Conclusion}

In conclusion, we have extended methods for statistical comparisons of single bit output ensemble quantum algorithms to classical probabilistic competitors to the Grover search algorithm. Our results indicate that the polarization required for the ensemble quantum algorithm to outperform known classical competitors scales as $N^{-1/4}$ where $N$ is the database size and the ensemble size is scaled so as to fix the failure probability of the classical algorithm. In absolute terms, the required polarization is modest compared to that required for an advantageous ensemble realization of the Deutsch-Jozsa algorithm.

\acknowledgments

Much of this work was carried out while the authors were in the Department of Physics and Astronomy at Bucknell University, Lewisburg, PA. We would like to thank that institution for the use of its resources and facilities as well as for support and encouragement from our colleagues there. 

\appendix


\section{Success probability for two qubits}
\label{app:twoqubitssuccess}

We aim to compute the probability with which qubit 1 returns the correct value $k_1$ times and qubit 2 returns the correct value $k_2$ times. Consider the joint measurement outcomes on each ensemble member; denote a typical outcome by $x_2x_1$ where the subscript indicates the qubit number. Let $l_0$ denote the number of times that $00$ occurs, $l_1$ the number of times that $01$ occurs, $l_2$ the number of times that $10$ occurs, etc \ldots. The probability with which this occurs is
\begin{equation}
  (l_0,l_1,l_2,l_3)! \left( \frac{1-\eps}{4} \right)^{l_0}
                    \left( \frac{1-\eps}{4} \right)^{l_1}
                    \left( \frac{1-\eps}{4} \right)^{l_2}
                    \left( \frac{1+3\eps}{4} \right)^{l_3}
\end{equation}
where the multinomial symbol is
\begin{equation}
	(l_0,l_1,l_2,l_3)! := \frac{(l_0+l_1+l_2+l_3)!}{l_0!\, l_1!\, l_2! \,l_3!}.
\end{equation}
Given values for $k_1$ and $k_2,$ various combinations of $l_0,l_1,l_2,l_3$ are possible; all are subject to the following constraints:
   \begin{subequations}
      \label{eq:firstconstraints}
      \begin{align}
	     l_0,l_1,l_2,l_3 & \geqslant 0  \label{eq:lpos} \\
		 l_0+l_1+l_2+l_3 & = M  \\
		 l_1 + l_3 & = k_1  \\  
		 l_2 + l_3 & = k_2.  
	  \end{align}
	\end{subequations}
Thus
   \begin{subequations}
    \label{eq:intermsofl3}
	\begin{align}
		    l_2 & = k_1 - l_3 \\
		    l_1 & = k_2 - l_3 \\
		    l_0 & = M -k_1 -k_2 + l_3
	\end{align}
\end{subequations}
and together with Eq.~\eqref{eq:lpos} these give
   \begin{subequations}
      \label{eq:secondconstraints}
      \begin{align}
	     l_3 & \geqslant k_1+k_2-M  \label{eq:l0l3} \\ 
	     k_1 & \geqslant l_3 \label{eq:l3k1} \\
	     k_2 & \geqslant l_3 \label{eq:l3k2} \\
	     l_3 & \geqslant 0 \label{eq:l3}
	  \end{align}
	\end{subequations}
Eqs.~\eqref{eq:l3k1} and~\eqref{eq:l3k2} are equivalent to
\begin{equation}
	l_3 \leqslant \min{(k_1,k_2)}.
	\label{eq:l3lower}
\end{equation}
while Eqs.~\eqref{eq:l0l3} and \eqref{eq:l3} are equivalent to
\begin{equation}
	l_3 \geqslant \max{(0,k_1+k_2-M)}. 
		\label{eq:l3upper}
\end{equation}
Any value of $l_3$ such that $\max{(0,k_1+k_2-M)} \leqslant l_3  \leqslant \min{(k_1,k_2)}$ together with Eqs.~\eqref{eq:intermsofl3}  yield $k_1$ and $k_2.$ Thus
\begin{widetext}
\begin{align}
      \Pr{(k_1,k_2)} = & \sum_{l_3=\max{(0,k_1+k_2-M)}}^{\min{(k_1,k_2)}} 
                          (M -k_1 -k_2 + l_3, k_1 - l_3, k_2 - l_3, l_3)! \nonumber \\
                       & \times 
                       \left( \frac{1-\eps}{4} \right)^{M -k_1 -k_2 + l_3}
                       \left( \frac{1-\eps}{4} \right)^{k_1-l_3}
                       \left( \frac{1-\eps}{4} \right)^{k_2-l_3}
                       \left( \frac{1+3\eps}{4} \right)^{l_3} \nonumber \\
                     = & \; \frac{1}{4^M}\, \sum_{l=\max{(0,k_1+k_2-M)}}^{\min{(k_1,k_2)}}
                       (M -k_1 -k_2 + l, k_1 - l, k_2 - l, l)! 
                       \left( 1-\eps \right)^{M-l}\,
                       \left( 1+3\eps \right)^{l}.               
	\end{align}
\end{widetext}
%


\section{Upper bound on failure probability}
\label{app:failurebounds}

The inequality
\begin{equation}
	\pfallq{\eps}{M}{N} \leqslant n\, \pfoneq{\eps}{M}{N}
\end{equation}
is clearly true for $n=1.$ We adopt the following notation for the $n$ qubit case: $\Pr{(s_n, \ldots, s_3, f_2,s_1)}$ is the probability that we succeed in identifying bit $1$ correctly, fail on bit $2$, succeed on bit $3$ and so on. For $n=2,$
   \begin{align}
	\pfallq{\eps}{M}{N} & = \Pr{(f_2,f_1)} + \Pr{(f_2,s_1)} + \Pr{(s_2,f_1)}
	                           \nonumber \\
	                       & = \Pr{(f_2,f_1)} + \Pr{(f_2,s_1)} + \Pr{(s_2,f_1)} 
	                           \nonumber \\
	                       & \phantom{=} + \Pr{(f_2,f_1)} - \Pr{(f_2,f_1)}.
	                           \nonumber 
   \end{align}
However, for bit $2$,
\begin{equation}
	\pfoneq{\eps}{M}{N} = \Pr{(f_2,f_1)} + \Pr{(f_2,s_1)}
\end{equation}
with a similar result for bit $2.$ Thus
   \begin{align}
	\pfallq{\eps}{M}{N} & = 2\, \pfoneq{\eps}{M}{N} - \Pr{(f_2,f_1)}
	                           \nonumber \\
	                       & \leqslant 2\, \pfoneq{\eps}{M}{N}  \nonumber 
   \end{align}
since $\Pr{(f_2,f_1)} \geqslant 0.$ The general result follows by induction on the number of qubits. For $n$ qubits,
 \begin{widetext}
 \begin{align}
	\pfallq{\eps}{M}{N}     & =  \Pr{(f_n, \textrm{any outcome on first $n-1$})} 
	                            + \Pr{(s_n,\textrm{fail in some way on first $n-1$})} 
	                           \nonumber \\
	                        & =  \pfoneq{\eps}{M}{N}
	                           + \Pr{(s_n,\textrm{fail in some way on first $n-1$})} 
	                           \nonumber \\
	                        &=  \pfoneq{\eps}{M}{N}
	                        + \Pr{(s_n,\textrm{fail in some way on first $n-1$})} 
	                           \nonumber \\
	                        & \phantom{=}
	                           + \Pr{(f_n,\textrm{fail in some way on first $n-1$})}
	                           - \Pr{(f_n,\textrm{fail in some way on first $n-1$})}
	                           \nonumber \\
	                        & \leqslant  \pfoneq{\eps}{M}{N}
	                                     + \Pr{(s_n,\textrm{fail in some way on first $n-1$})} 
	                           \nonumber \\
	                        & \phantom{=} + \Pr{(f_n,\textrm{fail in some way on first $n-1$})}
	                           \nonumber \\
	                        & \leqslant  \pfoneq{\eps}{M}{N} +  (n-1)\, \pfoneq{\eps}{M}{N}
	                           \nonumber \\
	                        & \leqslant  n\, \pfoneq{\eps}{M}{N}
   \end{align}
  \end{widetext}
which proves the result.


\begin{widetext}

\section{Perfect polarization bound for single bit failure probability}
\label{app:epsone}

The single bit failure probability is
\begin{align}
	\pfoneq{\eps}{M}{N} & = \frac{1}{2}\, \cbd{M}{\Mmin}{\effmn}{\effpn} 
	                        \nonumber \\
	                    & \phantom{=}+ \frac{1}{2} \cbd{M}{M- \Mmin+1}{\effmn}{\effpn}.
\label{eq:pfailone}
\end{align}
\end{widetext}
For $\eps=1$ and $q=q_\mathrm{std},$
\begin{equation}
  1-\frac{1}{N} \leqslant \lvert \alpha_q \rvert^2\, \leqslant 1
\end{equation}
giving
\begin{equation}
  1-\frac{1}{N-1} \leqslant  \epseff \leqslant 1. 
\end{equation}
Thus
\begin{align}
  \frac{1-\epseff}{2} & \leqslant \frac{1}{2(N-1)} 
                        \nonumber \\
  \frac{1+\epseff}{2} & \leqslant 1  \nonumber          
\end{align}
which implies that the first term in Eq~\eqref{eq:pfailone} satisfies
\begin{widetext}
\begin{align}
	 \frac{1}{2}\, \cbd{M}{\Mmin}{\effmn}{\effpn} & \leqslant  \frac{1}{2}\, 
	                                                        \sum_{k=\Mmin}^M 
	                                                        \frac{1}{2^k}\, 
	                                                        \frac{1}{(N-1)^k}
	                                                        \nonumber \\
	                                              & =  \frac{1}{2}\,
	                                                          \binom{M}{\Mmin}\,
	                                                          \frac{1}{[2(N-1)]^{\Mmin}}\,
	                                                         \nonumber \\
	                                              & \phantom{=} \quad
	                                                          \times
	                                                          \biggl\{ 1 + \frac{1}{2(N-1)}\, 
	                                                                       \binom{M}{\Mmin +1}/\binom{M}{\Mmin} 
	                                                                     + \ldots
	                                                                  \biggr\}
	                                                        \nonumber \\
	                                               & \leqslant \frac{1}{2}\,
	                                                          \binom{M}{\Mmin}\,
	                                                          \frac{1}{[2(N-1)]^{\Mmin}}\,
	                                                          \biggl\{ 1 + \frac{1}{2(N-1)} 
	                                                                     + \ldots
	                                                                  \biggr\}
	                                                         \nonumber \\
	                                               & \leqslant \frac{1}{2}\,
	                                                          \binom{M}{\Mmin}\,
	                                                          \frac{1}{[2(N-1)]^{\Mmin}}\,
	                                                          \frac{2(N-1)}{2N-3}
	                                                          \label{eq:pfonefirst}
\end{align}
\end{widetext}
since
\begin{equation}
	\sum_{k=0}^\infty \frac{1}{x^k} = \frac{x}{x-1}.
\end{equation}
Stirling's approximation gives~\cite{anderson05}
\begin{equation}
	\binom{M}{\Mmin} \approx 2^M \sqrt{\frac{2}{\pi M}}
\end{equation} 
for $M \gg 1$ and thus the first term in Eq~\eqref{eq:pfailone} is bounded by
%
\begin{align}
	& \frac{1}{2}\, \cbd{M}{\Mmin}{\effmn}{\effpn}  
	  \nonumber \\
	&	\leqslant \frac{1}{2}
	                                                          \sqrt{\frac{2}{\pi M}}\,
	                                                          \frac{2(N-1)}{2N-3}\,
	                                                          \left( \frac{2}{N-1} \right)^{M/2}.
\end{align}
A similar argument applies to the second term in Eq~\eqref{eq:pfailone}, giving the same bound. Thus 
\begin{equation}
	\pfoneq{1}{M}{N} \leqslant \sqrt{\frac{2}{\pi M}}\, 
	                              \frac{2(N-1)}{2N-3}\,
	                              \left( \frac{2}{N-1} \right)^{M/2}.
\end{equation}
%


\section{Gaussian approximation to the quantum failure probability}
\label{app:gaussapprox}

The single bit failure probability can be expressed exactly in terms of the incomplete beta function
\begin{equation}
  \begin{split}
	\pfoneq{\eps}{M}{N} = & \frac{1}{2}
	                        \bigl[ I_p(\Mmin,M-\Mmin+1)
	                              \\
	                              & \quad + I_p(M-\Mmin+1,\Mmin) 
	                             \bigr]
	                           \end{split}
	                       \label{eq:pfqbetaapp}
\end{equation}
where
\begin{equation}
	I_p(x,y) := \frac{\Gamma(x+y)}{\Gamma(x) \Gamma(y)}\, 
	            \int_0^p t^{x-1}(1-t)^{y-1},
	            \label{eq:betaapp}
\end{equation}
 $p=(1-\eps)/2,$ and 
\begin{equation}
	\Mmin = \left\lceil \frac{M+1}{2} \right\rceil.
\end{equation}
Substituting from Eq.~\eqref{eq:betaapp} into Eq.~\eqref{eq:pfqbetaapp} gives
\begin{widetext}
\begin{align}
 \pfoneq{\eps}{M}{N} & = \frac{1}{2}\,
 	                       \frac{\Gamma(M+1)}{\Gamma(\Mmin) \Gamma(M-\Mmin +1 )}\,
 	                       \int_0^p 
 	                              \left[ t^{\Mmin-1}(1-t)^{M-\Mmin} \right.
 	                              \nonumber \\
 	                    & = \phantom{\frac{1}{2}\,
 	                       \frac{\Gamma(M+1)}{\Gamma(\Mmin) \Gamma(M-\Mmin +1 )}\,
 	                       \int_0^p }
 	                                \left.
 	                                + t^{M-\Mmin}(1-t)^{\Mmin-1} \right] 
 	                               \mathrm{d} t. 
\end{align}
Redefining the variable of integration via $t^\prime:= 1-2t$ gives
\begin{align}
 \pfoneq{\eps}{M}{N} & = \frac{1}{2^{M+1}}\,
 	                       \frac{\Gamma(M+1)}{\Gamma(\Mmin) \Gamma(M-\Mmin +1 )}\,
 	                       \int_\eps^1 \left[ 
 	                             (1-t^\prime)^{\Mmin-1}(1+t^\prime)^{M-\Mmin} \right.
 	                       \nonumber \\
 	                   & \phantom{= \frac{1}{2^{M+1}}\,
 	                       \frac{\Gamma(M+1)}{\Gamma(\Mmin) \Gamma(M-\Mmin +1 )}\,
 	                       \int_\eps^1}
 	                             \left.
 	                             + (1-t^\prime)^{M-\Mmin}(1+t^\prime)^{\Mmin-1}
 	                             \right] \mathrm{d} t^\prime
 	                       \nonumber \\
 	                   & = \frac{1}{2^{M+1}}\,
 	                       \frac{\Gamma(M+1)}{\Gamma(\Mmin) \Gamma(M-\Mmin +1 )}\,
 	                       \left\{
 	                         \int_0^1 \left[ \ldots
 	                       \right] \mathrm{d} t^\prime 
 	                        - \int_0^\eps \left[ \ldots
 	                       \right] \mathrm{d} t^\prime 	  
 	                       \right\}.                    
\end{align}
However, 
\begin{align}
	     \pfoneq{0}{M}{N} & = \frac{1}{2^{M+1}}\,
 	                          \frac{\Gamma(M+1)}{\Gamma(\Mmin) \Gamma(M-\Mmin +1 )}\,
 	                       \int_0^1 \left[ \ldots
 	                       \right] \mathrm{d} t^\prime
 	                        = \frac{1}{2}
\end{align}
and thus
\begin{align}
 \pfoneq{\eps}{M}{N} & = \frac{1}{2} - \frac{1}{2^{M+1}}\,
 	                       \frac{\Gamma(M+1)}{\Gamma(\Mmin) \Gamma(M-\Mmin +1 )}\,
 	                       \int_0^\eps
 	                       \left[
 	                             (1-t^\prime)^{\Mmin-1}(1+t^\prime)^{M-\Mmin} \right.
 	                       \nonumber \\
 	                   & \phantom{= \frac{1}{2} - \frac{1}{2^{M+1}}\,
 	                       \frac{\Gamma(M+1)}{\Gamma(\Mmin) \Gamma(M-\Mmin +1 )}\,
 	                       \int_0^\eps} 
 	                       \left.    
 	                             + (1-t^\prime)^{M-\Mmin}(1+t^\prime)^{\Mmin-1}
 	                       \right] \mathrm{d} t^\prime.  
 	                    \label{eq:lastpfqoneexact}                  
\end{align}
\end{widetext}
The gamma function satisfies $\Gamma(n+1) = n!$ for integral $n,$ giving
\begin{equation}
	\frac{\Gamma(M+1)}{\Gamma(\Mmin) \Gamma(M-\Mmin +1 )} 
	= \frac{M!}{(M-\Mmin)!\,(\Mmin-1)!}.
\end{equation}
For $M \gg 1,$ $\Mmin \approx M/2$ and Stirling's approximation gives~\cite{anderson05}
\begin{equation}
	\frac{\Gamma(M+1)}{\Gamma(\Mmin) \Gamma(M-\Mmin +1 )} 
	\approxeq 2^M\,  \sqrt{\frac{M}{2\pi}}.
	\label{eq:agammapprox}
\end{equation}
Considering the expression of Eq.~\eqref{eq:lastpfqoneexact}, each of the exponents within the integral are approximately $M/2.$ Thus
\begin{equation}
 \pfoneq{\eps}{M}{N} \approxeq \frac{1}{2} - 
 	                       \sqrt{\frac{M}{2\pi}}
 	                       \int_0^\eps
 	                             (1-t^{\prime2})^{M/2}
 	                       \mathrm{d} t^\prime.  
 	                    \label{eq:firstpfqoneapprox}                  
\end{equation}
With $t^{\prime \prime}:= \sqrt{M} t^\prime,$ Eq.~\eqref{eq:firstpfqoneapprox} implies
\begin{equation}
	\pfoneq{\eps}{M}{N} \approxeq \frac{1}{2} - 
 	                       \frac{1}{\sqrt{2\pi}}
 	                       \int_0^{\eps \sqrt{M}}
 	                             \left(1-\frac{t^{\prime\prime 2}}{M}\right)^{M/2}
 	                       \mathrm{d} t^{\prime\prime}  
 	                    \label{eq:secondpfqoneapprox}     
\end{equation}
and for $M \gg 1$ the integrand approximates an exponential function, giving
\begin{equation}
	\pfoneq{\eps}{M}{N} \approxeq \frac{1}{2} - 
 	                       \frac{1}{\sqrt{2\pi}}
 	                       \int_0^{\eps \sqrt{M}}
 	                         e^{-t^2/2}
 	                       \mathrm{d} t.  
 	                    \label{eq:thirdpfqoneapprox}     
\end{equation}
%


\section{Necessary critical polarization vs $M$}
\label{app:necvsM}

We show that $\epscritone(M,N)$ is a monotonically increasing function of $M.$ In general $\epscritone$ satisfies
\begin{equation}
		\pfoneq{\epscritone(M,N)}{q}{N}{M} = 1- \frac{qM}{N}.
		\label{eq:neccritapp}
\end{equation}
For $M$ odd and any $\eps$ it can be shown~\cite{anderson05} that $\pfoneq{\eps}{M}{N} = \pfoneq{\eps}{M+1}{N}.$ The right side of Eq.~\eqref{eq:neccritapp} decreases as $M$ increases. Thus $\pfoneq{\epscritone(M+1,N)}{M+1}{N} < \pfoneq{\epscritone(M,N)}{M}{N}$ for odd $M$ and since $\pfoneq{\eps}{M}{N}$ decreases monotonically as $\eps$ increases, this implies that, for odd $M,$ $\epscritone(M+1,N) > \epscritone(M,N).$ It remains to consider $\epscritone(M+2,N)$ versus $\epscritone(M,N)$ for odd $M.$ Consider the ratio of the success probabilities,
\begin{equation}
	r(\eps,M,N):= \frac{\psoneq{\eps}{M}{N}}{\psallc{qM}{N}}
\end{equation}
where $\psallc{qM}{N} = qM/N$ is the probability with which the classical sequential search succeeds and $\psoneq{\eps}{M}{N} = 1 - \pfoneq{\eps}{M}{N}.$ We shall prove that $\Delta r(\eps,M,N) := r(\eps,M+2,N)- r(\eps,M,N)<0$ whenever $M$  is odd. Then $r(\epscritone(M,N),M,N)=1$ implies that $r(\epscritone(M,N),M+2,N) < r(\epscritone(M,N),M,N)=1.$ Thus $\epscritone(M+2,N)>\epscritone(M,N).$ Now, for $M$ odd~\cite{anderson05},
\begin{widetext}
\begin{align}
	\psoneq{\eps}{M}{N} & = \sum_{k=\frac{M+1}{2}}^M
	                           \binom{M}{k}
	                           \left( \frac{1+\eps}{2}\right)^k
	                           \left(\frac{1-\eps}{2}\right)^{M-k} 
	                           \nonumber \\
	                       & = I_p\left( \frac{M+1}{2}, \frac{M+1}{2}\right)
\end{align}
%
where $p = (1+ \eps)/2.$
In these terms
\begin{align}
	\Delta r(\eps,M,N) & = \frac{N}{q}
	   										\left[ \frac{1}{M+2}\, 
															I_p\left(\frac{M+3}{2},\frac{M+3}{2}\right)
											  \right.
											  \nonumber \\
										 & \phantom{= \frac{N}{q} [}
										    \left.
														  - \frac{1}{M}\, 
															I_p\left(\frac{M+1}{2},\frac{M+1}{2}\right) \right].
\end{align}
Then 
%
\begin{align}
	I_p\left(\frac{M+3}{2},\frac{M+3}{2}\right) & = \frac{\Gamma(M+3)}{\Gamma(\tfrac{M+1}{2}+1)\, \Gamma(\tfrac{M+1}{2}+1)}\,
	                                                \int_0^p 
	                                                t^{(M+1)/2} 
	                                                \left( 1-t\right)^{(M+1)/2}
	                                                \mathrm{d}t
	                                                \nonumber \\
	                                            & = \frac{(M+2)(M+1)}{[(M+1)/2]^2}\,
	                                                \frac{\Gamma(M+1)}{\Gamma(\tfrac{M+1}{2})\, \Gamma(\tfrac{M+1}{2})}
	                                                \int_0^p 
	                                                t^{(M+1)/2} 
	                                                \left( 1-t\right)^{(M+1)/2}
	                                                \mathrm{d}t
\end{align}
and thus
\begin{align}
	\Delta r(\eps,M,N) & = \frac{N}{q}\,
	                     \frac{\Gamma(M+1)}{[\Gamma(\tfrac{M+1}{2})]^2}
	                     \int_0^p
	                     \left\{ \frac{4}{M+1}\, 
	                             t^{(M+1)/2} 
	                             \left( 1-t\right)^{(M+1)/2} 
	                      - \frac{1}{M}
	                             t^{(M+1)/2} 
	                             \left( 1-t\right)^{(M+1)/2} 
	                     \right\}	
	                     \mathrm{d}t
	                     \nonumber \\
	                   & = \frac{N}{q}\,
	                     \frac{\Gamma(M+1)}{[\Gamma(\tfrac{M+1}{2})]^2}\,
	                     \int_0^p
	                     \left\{ \frac{4}{M+1}\, 
	                             t
	                             \left( 1-t\right)
	                             - \frac{1}{M}
	                      \right\}
	                      t^{(M-1)/2} 	
	                      \left( 1-t\right)^{(M-1)/2}
	                      \mathrm{d}t. 
	                     \label{eq:deltareq1}                  
\end{align}
\end{widetext}
For $0 \leqslant t \leqslant 1$ it is easily verified that $0 \leqslant t \left( 1-t\right) \leqslant 1/4.$ Thus the term in parenthesis within the integral of Eq.~\eqref{eq:deltareq1} is negative (except at the single point $t = 1/2$) and $\Delta r(\eps,M,N) < 0.$




\end{document}